
\magnification=1200
\baselineskip=12pt

\rightline{UR-1353$\ \ \ \ \ \ \ $}
\rightline{ER-40685-804}

\baselineskip=18pt
\medskip

\centerline{\bf COMMENT ON ``EQUIVALENCE OF SEVERAL}

\centerline{\bf CHERN-SIMONS MATTER MODELS"}

\vskip 1in
\centerline{by}

\vskip 1in

\centerline{C.R. Hagen}
\centerline{Department of Physics and Astronomy}
\centerline{University of Rochester}
\centerline{Rochester, NY 14627}

\vfill\eject

It has recently been claimed$^1$ that there is a complete equivalence
between three different models each of which consists of a matter field
minimally coupled to a Chern-Simons gauge field as well as to an
external vector field $C_\mu$.  These three cases consist of systems of
spins ${1 \over 2}$, 1, and ${3 \over 2}$ but each with a
 different coefficient
for the Chern-Simons term.  The object of this note is to
point out that such a claim is totally untenable.

It is simplest to begin by disposing of the spin-${3 \over 2}$ model.  As is
well known, a result of Johnson and Sudarshan$^2$ requires the matrix
multiplying the time derivative term in the Lagrangian to be singular
 in order to have a positive definite metric$^3$.
Since the matrices of ref. 1 are simply the non-singular spin-${3 \over 2}$
representation of angular momentum, one concludes that Eq. (3) of ref. 1
describes an indefinite metric theory albeit with twice as many degrees of
freedom as the single spin component models of Eqs. (1) and (2).
Alternatively stated, since no one has succeeded to date in consistently
quantizing the minimal spin-${3 \over 2}$ field coupled to a gauge field, it
is quite impossible to establish an equivalence to the relatively well
 understood
theories of Eqs. (1) and (2).

Next one turns to the equivalence claimed between the spinor and vector
theories.  The authors of ref. 1 discuss the current correlation
 functions at the one
loop level in support of their result.  It is known that
this quantity is finite
for all momenta
 $p$ in the spinor case but divergent for the vector field$^4$.  In
ref. 1 this complication is handled by
the remark that dimensional regularization$^5$ can eliminate the divergence at
$p=0$.  No additional momentum dependence is calculated and it is merely
hypothesized that all higher order corrections  do not affect the
conclusion.  In fact one can readily calculate the $O(p^2/M^2)$ corrections to
$\Pi_e$ and $\Pi_o$ using results of ref. 4.  The answer is that the
magnitude of the (finite) $O(p^2/M^2)$ correction to $\Pi_o$
 for the vector model is
exactly twice that of the spinor model, a result which directly contradicts
the equivalence claim of ref. 1.

Since the Aharonov-Bohm scattering problems for spin-${1 \over 2}^6$ and
spin one$^7$ have been extensively studied, it is also of interest to
make use of the fact that the results in these two cases are totally
dissimilar.  They can be realized in the context of ref. 1 by taking
 $\alpha \rightarrow \infty$ and $C_\mu$ to be the vector potential of a
single flux tube.  In the spinor case the solution is singular but
normalizable even when the Zeeman term is attractive while the vector model
has totally different properties.  Significantly, there exists no state of
lowest energy for the latter when the spin is oriented parallel to the
flux tube.

In summary then it is clear that the equivalence claims made in ref. 1 are
totally incompatible with well known results.

This work was supported in part by U.S. Department of Energy Grant No.
DE-FG-02-91ER40685.

\bigskip
\bigskip
\bigskip

\noindent {\bf References}

\medskip

\item{1.} W. Chen and C. Itoi, Phys. Rev. Lett. {\bf 72}, 2527 (1994).

\item{2.} K. Johnson and E.C.G. Sudarshan, Ann. Phys. (N.Y.) {\bf 13},
126 (1961).

\item{3.} It is straightforward to verify that the theorem applies also to
2 + 1 space.

\item{4.} C.R. Hagen, P. Panigrahi, and S. Ramaswamy, Phys. Rev. Lett.
{\bf 61}, 389 (1988).

\item{5.} This would seem to raise the question as to why dimensional
regularization should be used in this calculation since it is not
required for the path integral derivation.

\item{6.} C.R. Hagen, Phys. Rev. Lett. {\bf 64}, 503 (1990) and {\bf 64}
2347 (1990).

\item{7.} C.R. Hagen and S. Ramaswamy, Phys. Rev. {\bf D42}, 3524 (1990).

\end